\documentclass{rnaastex}

\begin{document}

\title{The Rise and Peak of the Luminous Type~IIn SN~2017hcc/ATLAS17lsn from
  ASAS-SN and Swift UVOT Data}

\author{J.~L.~Prieto}
\affiliation{N\'ucleo de Astronom\'ia, Facultad de Ingenier\'ia y
  Ciencias, Universidad Diego Portales, Ej\'ercito 441, Santiago,
  Chile} 

\author{Ping~Chen}
\affiliation{Kavli Institute for Astronomy and Astrophysics, Peking University, Yi He Yuan Road 5, Hai Dian District, Beijing 100871, China}

\author{Subo~Dong}
\affiliation{Kavli Institute for Astronomy and Astrophysics, Peking University, Yi He Yuan Road 5, Hai Dian District, Beijing 100871, China}

\author{B.~J.~Shappee}
\affiliation{Institute for Astronomy, University of Hawai'i, 2680 Woodlawn Drive, Honolulu, HI 96822, USA}

\author{M.~Seibert}
\affil{The Observatories of the Carnegie Institution for Science, 813 Santa Barbara St., Pasadena, CA 91101, USA}

\author{D.~Bersier}
\affil{Astrophysics Research Institute, Liverpool John Moores University, 146 Brownlow Hill, Liverpool L3 5RF, UK}

\author{T.~W.-S.~Holoien} \affil{The Observatories of the Carnegie Institution for Science, 813 Santa Barbara St., Pasadena, CA 91101, USA}

\author{C.~S.~Kochanek} 
\affil{Department of Astronomy, The Ohio State
  University, 140 W. 18th Avenue, Columbus, OH 43210, USA.} 

\author{K.~Z.~Stanek} \affil{Department of Astronomy, The Ohio State
  University, 140 W. 18th Avenue, Columbus, OH 43210, USA.}

\author{T.~A.~Thompson} \affil{Department of Astronomy, The Ohio State
  University, 140 W. 18th Avenue, Columbus, OH 43210, USA.}


\section{Introduction}

Type~IIn supernovae (SNe) are a heterogenous class of stellar
explosions associated with the deaths of massive stars surrounded by
dense circumstellar environments created by strong mass-loss
episodes \citep[e.g.,][]{smith14}. In these objects, the interaction between the
SN ejecta and a dense, H-rich circumstellar envelope dominates the
evolution of their light curves and spectra. 
  
SN~2017hcc/ATLAS17lsn was discovered as an optical transient on UT 2017 Oct. 2.38 by the
Asteroid Terrestrial-impact Last Alert System
\citep[ATLAS;][]{tonry11} at an orange filter magnitude of $o = 17.44$~mag (TNS~13920\footnote{\url{https://wis-tns.weizmann.ac.il/object/2017hcc/discovery-cert}}). Its
non-detection by ATLAS on 2017 Sep. 30.43 ($o \lesssim 19.04$~mag) showed that the transient
was caught early. We obtained a low resolution optical spectrum on Oct.
7.4 with FLOYDS mounted on the FTS~2m telescope at the Siding Spring Observatory that
showed the characteristics of a young Type~IIn SN at $z=0.0173$ (TNS~1284\footnote{\url{https://wis-tns.weizmann.ac.il/object/2017hcc/classification-cert}}).

\section{Observations and Analysis}

\begin{figure}[h!]
\begin{center}
\includegraphics[scale=1.2]{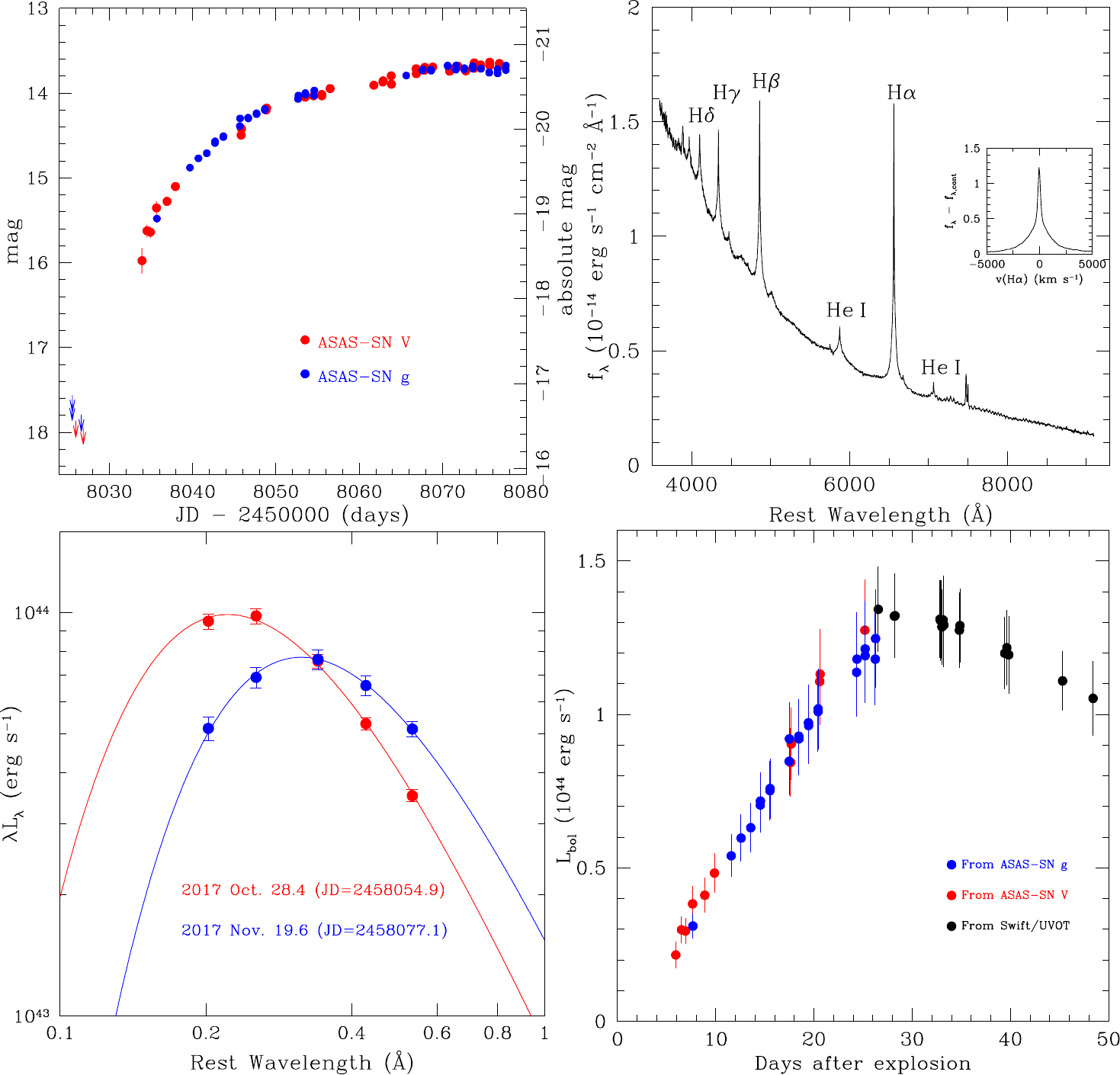}
\caption{{\it Top Left}: ASAS-SN $V$ (red) and $g$-band (blue) light curves of
  SN~2017hcc. {\it Top Right}: du~Pont spectrum obtained on UT
  2017 Oct. 20.08 showing a blue continuum and the
  characteristic features of a Type~IIn SN. {\it Bottom Left:} SED
  evolution obtained from observations at near-UV and optical
  wavelengths with the {\it
    Swift} UVOT. The lines show blackbody model fits with $\rm
  T_{bb}=16539 \pm 343$~K and $\rm R_{bb} = (1.59 \pm 0.07) \times
  10^{15}$~cm on UT 2017 Oct.~28.4 (red) and  $T_{bb}=11657 \pm 238$~K
  and $\rm R_{bb} = (2.83 \pm 0.08) \times
  10^{15}$~cm on UT 2017 Nov.~19.6 (blue).  {\it Bottom Right:} Evolution of
  the bolometric luminosity as a function of time. For
  the early rise (blue and red points) of the light curve we assume a blackbody SED with 
  the temperature fixed to $\rm T_{bb}=16539$~K obtained from the earliest
  {Swift} UVOT photometry to estimate a bolometric correction to the
  ASAS-SN $V$ and $g$-band photometry. For the later part of the light
  curve (black points), we derive the bolometric luminosity directly from 
  blackbody fits to the {\it Swift} UVOT photometry. \label{fig1}}
\end{center}
\end{figure}

We have been obtaining observations of SN~2017hcc as part of the All-Sky
Automated Survey for Supernovae \citep[ASAS-SN;][]{shappee14} using
the quadruple 14-cm ``Brutus'' telescopes in Haleakala, Hawaii, and the
recently installed quadruple 14-cm ``Bohdan Paczynski'' telescopes in CTIO,
Chile. The ASAS-SN images are reduced by an automated difference
imaging pipeline and the magnitudes of sources are calibrated using
standard $V$ (``Brutus'') or SDSS $g$-band (``Paczynski'')
photometry of local standard stars from the APASS DR9
\citep{henden12}. The SN was undetected by ASAS-SN down to
$V/g \lesssim 17-18$~mag on 2017 Sep. 29-30 and first detected rising
at $V=15.98 \pm 0.15$ on Oct. 7.4. Now it has reached peak
magnitude at $V\simeq 13.7$~mag. We show the current ASAS-SN light curve in Fig.~\ref{fig1}. 

We triggered {\it Swift} ToO observations that started
on Oct. 28.4. The near-UV and optical colors of SN~2017hcc are
consistent with a hot (but cooling) blackbody temperature with $W1-U
\simeq -0.4$~mag on Oct. 28.4 and $W1-U \simeq 0.0$~mag on Nov. 19.6. 
The SN is undetected in X-rays with the {\it Swift} XRT
\citep{chandra17} and shows very strong continuum polarization at
optical wavelengths \citep{mauerhan17}.
We obtained new spectra on 2017 Oct. 20-21
with WFCCD on the du~Pont 2.5m telescope at Las Campanas
Observatory. The spectra are characterized by a strong blue continuum
with Balmer and He~I emission lines (see top right panel of 
Fig.~\ref{fig1}), with line profiles characteristic of Type~IIn. We
measure a better constrained redshift of $z=0.0168$ ($D\simeq 73$~Mpc).
 
We used a blackbody function to fit the spectral energy distribution
(SED) of the SN using the {\it Swift} data and correcting for
Galactic extinction (bottom left panel of Fig.~\ref{fig1}). The fits show a hot, decreasing
blackbody temperature and an increasing blackbody radius, consistent
with early observations of SNe. Assuming a fixed temperature
for the early rise of the light curve where there is only ASAS-SN
data, we obtain the bolometric
light curve of SN~2017hcc shown in the bottom right panel of Fig.~\ref{fig1}. The
bolometric luminosity at peak $\rm L_{bol,peak}= (1.34\pm 0.14)\times
10^{44}$~erg~s$^{-1}$ ($\rm M_{V,peak} \simeq -20.7$~mag), makes
SN~2017hcc one of the most luminous Type~IIn SNe known 
\citep[e.g.,][]{smith07,ofek14,jencson16}. From the bolometric light
curve we constrain the risetime to be $\sim 27$~days and the total
radiated energy of the event to date is $\sim 4\times 10^{50}$~erg.

\acknowledgments

We thank Dr.~Brad~Cenko for approving the {\it Swift} ToO
observations. We thank Las Cumbres Observatory and its staff for their
continued support of ASAS-SN. ASAS-SN is funded in part by the Gordon
and Betty Moore Foundation through grant GBMF5490 to the Ohio State
University, NSF grant AST-1515927, the Mt. Cuba Astronomical
Foundation, the Center for Cosmology and AstroParticle Physics 
(CCAPP) at OSU, and the Chinese Academy of Sciences South America 
Center for Astronomy (CASSACA).


\begin{thebibliography}{}

\bibitem[Chandra et al.(2017)]{chandra17} Chandra, P., et al.\ 2017, ATel, 10936  
\bibitem[Henden et al.(2012)]{henden12} Henden, A.~A., et al.\ 2012, JAAVSO, 40, 430 
\bibitem[Jencson et al.(2016)]{jencson16} Jencson, J.~E., et al.\ 2016, \mnras, 456, 2622 
\bibitem[Mauerhan et al.(2017)]{mauerhan17} Mauerhan, J.~C., et al.\
  2017, ATel, 10911
\bibitem[Ofek et al.(2014)]{ofek14} Ofek, E.~O., et al.\ 2014, \apj, 788, 154 
\bibitem[Shappee et al.(2014)]{shappee14} Shappee, B.~J., et al.\ 2014, \apj, 788, 48 
\bibitem[Smith et al.(2007)]{smith07} Smith, N., et al.\ 2007, \apj, 666, 1116 
\bibitem[Smith(2014)]{smith14} Smith, N.\ 2014, \araa, 52, 487 
\bibitem[Tonry(2011)]{tonry11} Tonry, J.~L.\ 2011, \pasp, 123, 58 

\end{thebibliography}
\end{document}